\documentclass[]{interact}

\usepackage{epstopdf}% To incorporate .eps illustrations using PDFLaTeX, etc.
\usepackage[caption=false]{subfig}% Support for small, `sub' figures and tables
\usepackage[longnamesfirst,sort]{natbib}% Citation support using natbib.sty
\bibpunct[, ]{(}{)}{;}{a}{,}{,}% Citation support using natbib.sty
\usepackage{makecell}
\usepackage{xcolor}
\usepackage{url}

\theoremstyle{plain}% Theorem-like structures provided by amsthm.sty

\theoremstyle{definition}

\theoremstyle{remark}

\begin{document}

\title{\texttt{ChauBoxplot} and \texttt{AdaptiveBoxplot}: Two R packages for boxplot-based outlier detection}

\author{
\name{Tiejun Tong\textsuperscript{a},
Hongmei Lin\textsuperscript{b}\thanks{CONTACT Hongmei Lin. Email: hmlin@suibe.edu.cn},
Bowen Gang\textsuperscript{c} and
Riquan Zhang\textsuperscript{b}}
\affil{\textsuperscript{a}Department of Mathematics, Hong Kong Baptist University, Hong Kong, People's Republic of China;
       \textsuperscript{b}School of Statistics and Data Science, Shanghai University of International Business and Economics, Shanghai, People's Republic of China;
       \textsuperscript{c}Department of Statistics and Data Science, Fudan University, Shanghai, People's Republic of China}
}

\maketitle

\begin{abstract}
Tukey's boxplot is widely used for outlier detection; however, its classic fixed-fence rule tends to flag an excessive number of outliers as the sample size grows.
To address this, we introduce two new R packages, \texttt{ChauBoxplot} and \texttt{AdaptiveBoxplot}, which implement more robust and statistically principled outlier detection methods.
We illustrate their advantages and practical implications through comprehensive simulation studies and a real-world analysis of provincial university admission rates from China's National College Entrance Examination.
Based on these findings, we provide practical guidance to help practitioners select appropriate boxplot methods, achieving a balance between interpretability and statistical reliability.
\end{abstract}

\begin{keywords}
Box-and-whisker plot; Chauvenet's criterion; Chauvenet-type boxplot; fence coefficient; outlier detection; sample size
\end{keywords}

\section{Introduction}

The box-and-whisker plot, first introduced by \citet{Tukey1977bk}, is a fundamental and widely used tool in exploratory data analysis.
It provides a concise summary of key distributional features, including the median, quartiles, and range, and visually identifies potential outliers.
Due to its non-parametric nature and ease of interpretation, the boxplot is commonly applied in fields such as biomedicine, economics, engineering, psychology, and social sciences, as well as in quality control and educational research.
A central function of the boxplot is outlier detection, which is performed by defining fences at fixed distances from the first quartile ($Q_1$) and the third quartile ($Q_3$), and flagging any observations outside these fences as potential outliers.
Tukey's original boxplot sets the lower and upper fences at
\begin{eqnarray*}
{\rm LF} = Q_1 - k \times \mathrm{IQR} ~~~~~{\rm and}~~~~~ {\rm UF} = Q_3 + k \times \mathrm{IQR},  \label{Tukey.fence}
\end{eqnarray*}
where $\mathrm{IQR} = Q_3 - Q_1$ is the interquartile range, and $k=1.5$ is the standard fence coefficient.
Although this outlier detection rule is simple and widely used, it lacks a rigorous theoretical foundation \citep{Hoaglin1986,Hoaglin1987,Frigge1989,Sim2005,Hofmann2017}.
Specifically, applying a fixed value of $k$ disregards the sample size and distributional characteristics, often leading to an excessive number of false positive outlier flags in larger samples.

To address the limitation imposed by a fixed fence coefficient, \citet{LinH2026} recently introduced the Chauvenet-type boxplot, which provides exact control of the outside rate per observation.
For normally distributed data, combining Tukey's boxplot with Chauvenet's criterion \citep{Chauvenet1863} yields a new fence coefficient, given by
\begin{eqnarray}
k_n^{\rm Chau} = \frac{\Phi^{-1}(1 - 0.25/n)}{1.35} - 0.5,    \label{Chauvenet.fence}
\end{eqnarray}
where $\Phi^{-1}$ denotes the standard normal quantile function and $n$ is the size of the data.
Moreover, \citet{GangB2026} reframed boxplot-based outlier detection as a multiple testing problem, treating each flagged observation as a statistical test against the majority of the data.
Their methods provide precise control over the family-wise error rate, the per-family error rate, or the false discovery rate.
These methodological advances are now available in statistical software, with two new R packages—\texttt{ChauBoxplot} and \texttt{AdaptiveBoxplot}—offering convenient tools for practical implementation.
However, the increased variety of options introduces new challenges, such as selecting suitable methods, interpreting error rates appropriately, and balancing simplicity with statistical rigor.

In this short paper, we present a focused review of two representative R packages used for boxplot-based outlier detection.
Section 2 introduces \texttt{ChauBoxplot}, which implements the Chauvenet-type boxplot to provide precise control over the outside rate for each observation and discusses its statistical foundation and practical applications.
Section 3 presents \texttt{AdaptiveBoxplot}, which utilizes adaptive multiple testing procedures to enhance the accuracy and optimizes threshold selection in outlier detection.
Section 4 presents comprehensive simulation studies comparing classical and modern boxplot approaches across a range of sample sizes.
Section 5 presents a real-world application by analyzing provincial university admission rates from China's National College Entrance Examination to illustrate the implications of boxplot-based outlier detection in practice.
Finally, Section 6 offers practical guidance on selecting boxplot-based outlier detection techniques, summarizing the strengths and limitations of each method.

\section{The \texttt{ChauBoxplot} package}

The Chauvenet-type boxplot introduces a novel fence coefficient that, on average, excludes half an observation from normally distributed data regardless of the sample size.
Instead of using Tukey's fixed fence coefficient, it employs the sample-size-adjusted alternative in Equation (\ref{Chauvenet.fence}) according to Chauvenet's criterion.
This adjustment integrates seamlessly into the standard boxplot construction, offering enhanced statistical reliability without sacrificing visual intuition or workflow convenience.
For further technical details and the theoretical foundation of the Chauvenet-type coefficient, readers may refer to \citet{LinH2026}.

The \texttt{ChauBoxplot} package provides two main user interfaces for creating the Chauvenet-type boxplot.
The \texttt{chau\underline{~}boxplot()} function acts as a direct drop-in replacement for base R's \texttt{boxplot()}, with the only difference being the default $k=1.5$ replaced by $k_n^{\rm Chau} = \Phi^{-1}(1 - 0.25/n)/1.35 - 0.5$.
For users within the \texttt{ggplot2} ecosystem, the \texttt{geom\underline{~}chau\underline{~}boxplot()} function can be used similarly to \texttt{geom\underline{~}boxplot()}.
It provides a dedicated grammar of graphics layer that supports grouped, faceted, and highly customizable plots, making it well-suited for advanced data analysis.
Specifically, the new boxplot can be generated using the following syntax:

\begin{verbatim}
		library(ChauBoxplot)

		# For base R:
		chau_boxplot()

		# For the ggplot2 ecosystem:
		ggplot() + geom_chau_boxplot()
\end{verbatim}
This compatibility enables researchers to adopt the new boxplot method with minimal changes to their existing code or analysis workflows.

The \texttt{ChauBoxplot} package offers comprehensive documentation, including a reference manual, a README file, and a set of vignettes that demonstrate usage with both synthetic and real-world datasets.
These resources include worked examples from our simulation studies and the analysis of provincial university admission rates presented in Sections 4 and 5, guiding users through both fundamental functions and advanced features.
Users can explore comparisons between Chauvenet's criterion and Tukey's classic rule, and learn how \texttt{ChauBoxplot} provides advantages in handling large samples or non-normal data.
The package is available on CRAN for all major R platforms under the GPL-3 license, and is actively maintained by the original authors to ensure it remains aligned with theoretical developments and evolving user needs.
By combining statistically sound, sample-size-adjusted outlier detection with an intuitive interface and thorough support materials, \texttt{ChauBoxplot} empowers researchers and analysts to conduct modern, reproducible data analysis across a broad range of scientific disciplines.

\section{The \texttt{AdaptiveBoxplot} package}

The unified framework by \citet{GangB2026} extends the methodological advances of \citet{LinH2026} by systematically reframing outlier detection as a multiple hypothesis testing problem.
From this perspective, Tukey's classic boxplot corresponds to an unadjusted testing procedure with a fixed per-comparison error rate, while the Chauvenet-type boxplot proposed by \citet{LinH2026} effectively controls the per-family error rate (PFER).
Building on this insight, \citet{GangB2026} developed a generalized pipeline to construct boxplots that control more advanced error metrics, including the family-wise error rate (FWER) and the false discovery rate (FDR). This pipeline proceeds in four steps:
\begin{enumerate}
	\item[1)] {\it \bf Robust Estimation:} Parameters of the null distribution are estimated using robust measures, typically the sample quartiles.
	\item[2)] {\it \bf $\boldsymbol{p}$-value Calculation:} Each observation is converted into a $p$-value based on the estimated null distribution.
    \item[3)] {\it \bf Multiple Testing Adjustment:} A desired error rate (e.g., FWER or FDR) and a control target $\alpha$ are first selected. Subsequently, a suitable multiple testing procedure is applied to the set of $p$-values to determine an adjusted significance threshold.
	\item[4)] {\it\bf Fence Construction:} The adjusted threshold is transformed back onto the original data scale to define the boxplot fences.
\end{enumerate}
This approach allows for the creation of fences that are adaptive not only to the sample size but also to the distributional characteristics of the data.
Control of the FWER ensures a conservative bound that is ideal for strict error avoidance, while control of the FDR offers a powerful alternative that scales effectively with data complexity.

The \texttt{AdaptiveBoxplot} package implements this pipeline through two primary functions.
The \texttt{holm\underline{~}boxplot()} function utilizes Holm's step-down procedure \citep{holm1979simple} to control the FWER.
This function is particularly well-suited for applications requiring strict conservatism, where avoiding even a single false positive is a priority.
In contrast, the \texttt{bh\underline{~}boxplot()} function applies the Benjamini-Hochberg procedure \citep{benjamini1995controlling} to control the FDR.
This method offers a more powerful approach for large-scale or complex datasets, as it adaptively relaxes the fences when strong evidence of outliers is present, thereby maintaining high sensitivity without succumbing to the inflation of false discoveries typical of fixed-threshold methods.

The package interface is designed for simplicity and consistency with base R graphics. Users can generate error-controlled boxplots for both univariate vectors and grouped data frames using the following syntax:

\begin{verbatim}
		library(AdaptiveBoxplot)
	
		# Controlling the FWER at level 0.05:
		holm_boxplot(data, alpha = 0.05)

		# Controlling the FDR at level 0.05:
		bh_boxplot(data, alpha = 0.05)
\end{verbatim}
Apart from the inclusion of the \texttt{alpha} argument, which specifies the target error rate, the usage of both \texttt{holm\underline{~}boxplot()} and \texttt{bh\underline{~}boxplot()} is essentially identical to that of the base function \texttt{boxplot()}.

The \texttt{AdaptiveBoxplot} package is freely available on CRAN under the GPL-3 license, supported by a concise reference manual that illustrates the function usage.
As an actively maintained project, it ensures that these advanced, multiple-testing-based boxplots remain accessible to the R community.
By embedding rigorous statistical inference into a familiar graphical tool, \texttt{AdaptiveBoxplot} facilitates more reproducible and defensible outlier detection.
The package's methods are demonstrated in the simulation study presented in Section 4, highlighting their practical advantages in controlling error rates for boxplot-based outlier detection.
For more details on the unified $p$-value pipeline, readers may refer to \citet{GangB2026}.

\section{Simulated data example}

To illustrate the practical differences among the boxplot-based outlier detection methods, we conducted a simulation study comparing four representative approaches: Tukey’s boxplot, the Chauvenet-type boxplot, the FWER (Holm) boxplot, and the FDR (BH) boxplot.
For each setting, we generated a dataset of size $n$, consisting of $n-3$ independent observations from the standard normal distribution $\mathcal{N}(0, 1)$ and three outliers sampled from
$\mathcal{N}(5, 0.5^2)$.
The sample sizes considered were $n = 50$, $500$, $5000$ and $50000$, respectively, allowing us to assess the performance across both small and large datasets.
The data generation and analysis were implemented as follows:

\begin{verbatim}
		library(ChauBoxplot)
		library(AdaptiveBoxplot)

		n <- c(50, 500, 5000, 50000)
		m <- 3
		outliers <- rnorm(m, 5, 0.5)

		par(mfrow = c(4,4))
		oldpar = par(mar = c(1, 3.5, 2, 0.5), mgp = c(2, 0.5, 0))

		for (i in 1:length(n)){
	  	x <- c(rnorm(n[i]-m), outliers)
	  	boxplot(x, main = "Tukey's boxplot", ylab=paste("n =", n[i]))
	  	chau_boxplot(x)
	  	holm_boxplot(x, alpha = 0.05)
	  	bh_boxplot(x, alpha = 0.05)
		}
\end{verbatim}

\begin{figure}
\center{
\includegraphics[width=14cm, angle=0]{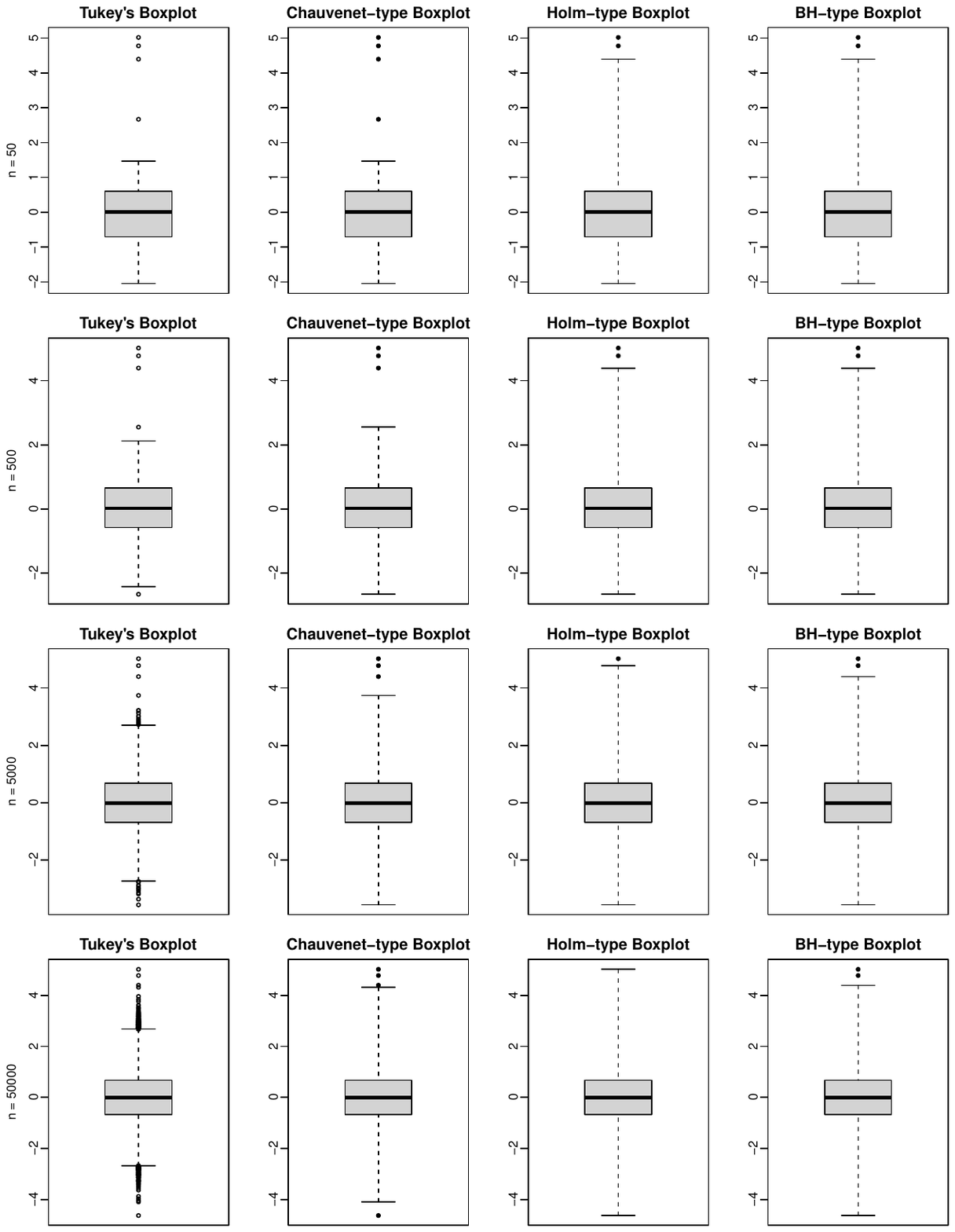}
\caption{
Comparison of boxplot-based outlier detection methods for normal data with three contaminated outliers.
Each panel displays the results for one method (Tukey, Chauvenet-type, FWER [Holm] with rate 0.05, or FDR [BH] with rate 0.05) at $n = 50$, $500$, $5000$ and $50000$, respectively.}
\label{figure1}
}
\end{figure}

Figure \ref{figure1} offers a comparative overview of the four boxplot-based outlier detection methods across a range of sample sizes.
Each row represents a different sample size, and each column depicts a distinct method, with detected outliers shown as points beyond the whiskers.
The three artificially introduced outliers are clearly visible in most panels.
As the sample size increases, the classical boxplot with $k=1.5$ tends to misclassify an increasing number of central observations as outliers, especially for $n=5000$ and $50000$, where many non-outlying values are incorrectly flagged.
In contrast, the Chauvenet-type, FWER (Holm), and FDR (BH) boxplots only flag very few observations in total, demonstrating their robust performance regardless of the sample size.
Finally, we note that Figure \ref{figure1} is fully reproducible: by setting \texttt{set.seed(1000)} for random number generation, the provided R code can be run directly to obtain the figure as shown.

These results underscore the advantages of sample-size-adjusted and error-controlled boxplots for reliable outlier detection in moderate to large datasets.
In contrast, the classical Tukey approach can lead to misleading conclusions when applied to large samples.
Our simulation study demonstrates that the adaptive strategies implemented in the \texttt{ChauBoxplot} and \texttt{AdaptiveBoxplot} packages offer principled control of error rates and maintain robust outlier identification across a wide range of sample sizes. These findings highlight the practical value of using statistically grounded methods for reproducible and defensible outlier detection.

\section{Real data example}

To further illustrate the application of boxplot-based outlier detection, we analyzed provincial admission rates from the 2024 National College Entrance Examination (Gaokao) in China.
The Gaokao is among the world's largest standardized tests, with a record 13.53 million participants in 2024.
In this analysis, we focused on admission rates to two groups of elite universities: the ``Project 985'', which comprises 39 institutions, and the ``Project 211'', which encompasses 115 universities.
These measures are widely used as indicators of regional educational opportunity.
Table \ref{Table1} presents the number of examinees and admission rates by province, ordered by the Project 985 admission rate.
Municipalities such as Tianjin, Beijing, and Shanghai have the highest Project 985 rates, reflecting concentrated resources, while provinces such as Jiangxi, Guizhou, and Anhui have among the lowest, underscoring regional disparities.

\begin{table}[htbp]
\caption{Provincial summary of Gaokao 2024 examinee counts, admissions to the Project 985 and Project 211 universities, and associated admission rates.
The table is sorted in descending order by the Project 985 admission rate. Admission rates denote the percentage of total examinees in each province admitted to the respective university category.  \\}
\centering
\begin{tabular}{c|c|c|c|c|c}
\hline
~Province~ & \makecell{Total \\ Examinees} & \makecell{985 \\ Admissions} &~~~ \makecell{985 \\ Rate}~~~ & \makecell{211 \\ Admissions} &~~~ \makecell{211 \\ Rate}~~~ \\
\hline
Tianjin        & 74000    & 4410    & 6.0\%  & 9588    & 13.0\% \\
Beijing        & 68000    & 3604    & 5.3\%  & 11319   & 16.7\% \\
Shanghai       & 58000    & 2541    & 4.4\%  & 6520    & 11.2\% \\
Jilin          & 132000   & 4230    & 3.2\%  & 10575   & 8.0\% \\
Qinghai        & 59000    & 1579    & 2.7\%  & 6159    & 10.4\% \\
Ningxia        & 82000    & 1686    & 2.1\%  & 6306    & 7.7\% \\
Shaanxi        & 350000   & 6983    & 2.0\%  & 18824   & 5.4\% \\
Liaoning       & 203000   & 4129    & 2.0\%  & 10593   & 5.2\% \\
Zhejiang       & 405000   & 7106    & 1.8\%  & 15905   & 3.9\% \\
Chongqing      & 350000   & 6198    & 1.8\%  & 15939   & 4.6\% \\
Hubei          & 520000   & 9513    & 1.8\%  & 23782   & 4.6\% \\
Shanxi         & 354000   & 6019    & 1.7\%  & 18690   & 5.3\% \\
Shandong       & 1000000  & 16544   & 1.7\%  & 35344   & 3.5\% \\
Fujian         & 247000   & 4098    & 1.7\%  & 11066   & 4.5\% \\
Jiangsu        & 480000   & 7633    & 1.6\%  & 26333   & 5.5\% \\
Heilongjiang   & 211000   & 3422    & 1.6\%  & 11462   & 5.4\% \\
Xinjiang       & 230000   & 3516    & 1.5\%  & 16130   & 7.0\% \\
Hainan         & 73000    & 1125    & 1.5\%  & 3849    & 5.3\% \\
Guangxi        & 480000   & 6768    & 1.4\%  & 25944   & 5.4\% \\
Guangdong      & 760000   & 10558   & 1.4\%  & 23096   & 3.0\% \\
Henan          & 1360000  & 18471   & 1.4\%  & 54182   & 4.0\% \\
Gansu          & 260000   & 3483    & 1.3\%  & 8126    & 3.1\% \\
Sichuan        & 830000   & 10857   & 1.3\%  & 31847   & 3.8\% \\
Xizang         & 39000    & 511     & 1.3\%  & 3850    & 9.9\% \\
Hunan          & 700000   & 9236    & 1.3\%  & 23397   & 3.3\% \\
Yunnan         & 406000   & 5040    & 1.2\%  & 16201   & 4.0\% \\
Hebei          & 880000   & 10143   & 1.2\%  & 32768   & 3.7\% \\
Inner Mongolia & 230000   & 2609    & 1.1\%  & 10434   & 4.5\% \\
Anhui          & 670000   & 7344    & 1.1\%  & 26552   & 4.0\% \\
Guizhou        & 506000   & 5392    & 1.1\%  & 23365   & 4.6\% \\
Jiangxi        & 650000   & 6430    & 1.0\%  & 24647   & 3.8\% \\
\hline
\end{tabular}\label{Table1}
\end{table}

\begin{figure}
\center{
\includegraphics[width=14cm, angle=0]{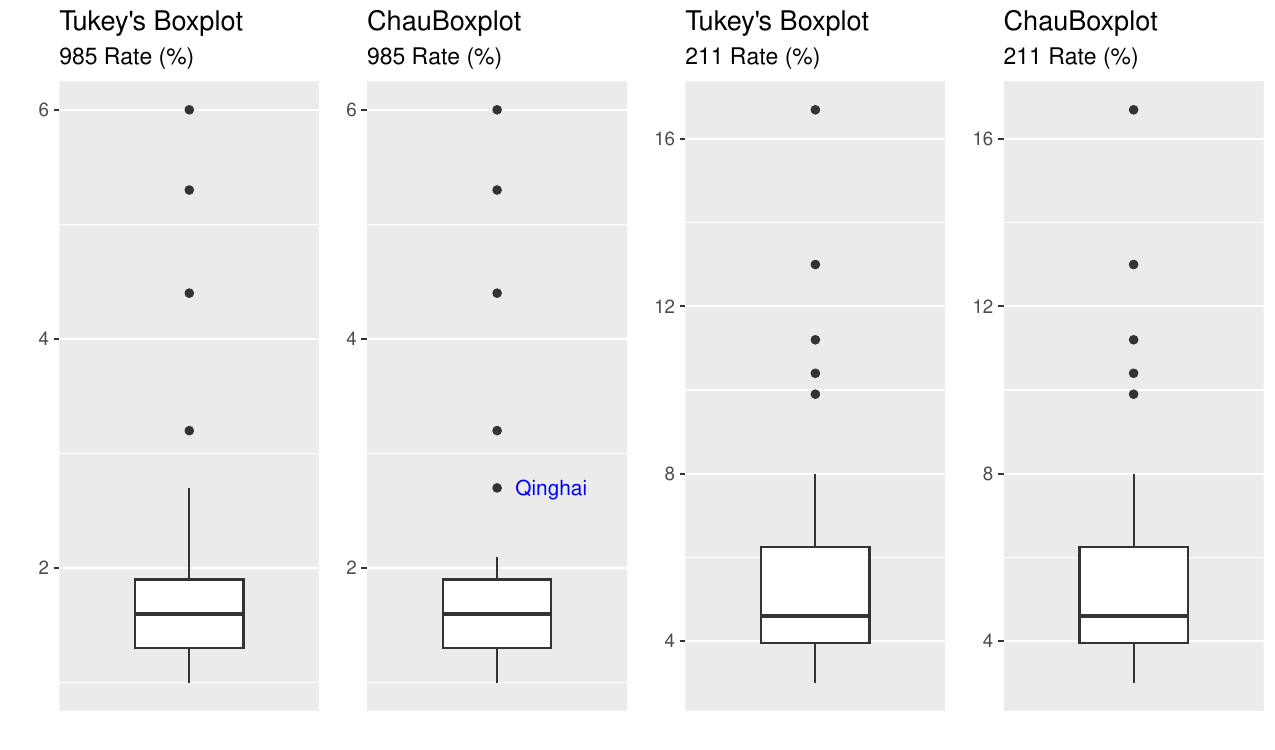}
\caption{Tukey's boxplot and the Chauvenet-type boxplot of the Gaokao admission rates in 2024. The left two panels are for the Project 985 university admission rates, while the right two panels are for the Project 211 university admissions rates.}
\label{figure2}
}
\end{figure}

More specifically, we use the admission rates for Project 985 and Project 211 universities from \url{https://www.gk100.com/read_12602336.htm} for this analysis.
All visualizations are produced with \texttt{ggplot2}, as Section 4 already covers base R graphics.
Because \texttt{AdaptiveBoxplot} supports only base R, this example focuses on comparing Tukey's classical boxplot with the sample-size-adjusted Chauvenet-type boxplot.
The complete analysis code, including scripts to reproduce Figure \ref{figure2} for both Project 985 and Project 211 admission rates, is available at \url{https://github.com/tiejuntong/ChauBoxplot}, ensuring the analysis is entirely reproducible.

\vskip 12pt
\begin{verbatim}
		library(ggplot2)
		library(ChauBoxplot)

		province <- c("Tianjin", ..., "Jiangxi")
		rate.985 <- c(6.0, ..., 1.0)
		gaokao <- data.frame(province, rate.985, rate.211)

		theme_box <- theme(legend.position = "none",
  		plot.margin = unit(c(0, 0, 0, 0), "inches"))

		ggplot(gaokao, aes(x = "", y = rate.985)) +
		  geom_boxplot(width = 0.5) +
		  scale_x_discrete(breaks = NULL) + theme_box +
		  labs(title = "Tukey's Boxplot", subtitle = "985 Rate (%)")

		ggplot(gaokao, aes(x = "", y = rate.985)) +
		  geom_chau_boxplot(width = 0.5) +
		  geom_text(data = subset(gaokao, province == "Qinghai"),
            mapping = aes(x = "", y = rate.985, label = province),
            nudge_x = 0.25, size = 4.5, color = "blue", vjust = 0.5) +
		  scale_x_discrete(breaks = NULL) + theme_box +
		  labs(title = "ChauBoxplot", subtitle = "985 Rate (%)")
\end{verbatim}

For the Project 985 admission rates, the first quartile ($Q_1$) is 1.3\% and the third quartile ($Q_3$) is 1.9\%, giving an interquartile range (IQR) of 0.6\%.
Tukey's boxplot flags observations above the upper fence, calculated as $Q_3 + 1.5 \times \mathrm{IQR} = 2.8\%$, identifying Tianjin (6.0\%), Beijing (5.3\%), Shanghai (4.4\%), and Jilin (3.2\%) as outliers.
In contrast, the Chauvenet-type boxplot uses a sample-size-adjusted coefficient, $k_n^{\rm Chau} = {\Phi^{-1}(1 - 0.25/n)}/{1.35} - 0.5 = 1.28$ with $n = 31$, yielding an upper fence of $Q_3 + 1.28 \times \mathrm{IQR} = 2.67\%$.
This approach flags the same high-rate provinces and, due to the slightly lower threshold, also identifies Qinghai (2.7\%) as a borderline outlier.
These results demonstrate that the adaptive Chauvenet-type boxplot is more sensitive to moderate but meaningful deviations, providing a nuanced view of regional disparities in educational opportunity.

For the Project 211 admission rates, both Tukey's boxplot and the Chauvenet-type boxplot identified Beijing (16.7\%), Tianjin (13.0\%), Shanghai (11.2\%), Qinghai (10.4\%), and Xizang (9.9\%) as outliers. These provinces are characterized by exceptionally high admission rates to the Project 211 universities, reflecting a strong concentration of educational resources in provincial-level municipalities as well as distinct regional patterns in Qinghai and Xizang.
Overall, these results show that both traditional and adaptive boxplot methods consistently highlight provinces with significantly greater access to elite higher education, while the Chauvenet-type approach offers additional sensitivity to borderline cases.

\section{Conclusion and practical guidance}

This paper reviewed recent advances in boxplot-based outlier detection, focusing on the \texttt{ChauBoxplot} and \texttt{AdaptiveBoxplot} R packages.
Results from simulations and real-world data demonstrate that Tukey's classic boxplot, with its fixed-fence rule, increasingly misclassifies central observations as outliers as the sample size grows,   while the sample-size-adjusted and error-controlled approaches, namely the Chauvenet-type, Holm-type, and BH-type boxplots, offer substantially improved reliability.
The Chauvenet-type boxplot provides explicit control over the expected number of flagged points, the Holm-type approach controls the probability of at least one false outlier, and the BH-type method adapts flexibly while bounding the false discovery rate. Table~\ref{tab:boxplot-summary} provides a concise summary of the error control and practical characteristics of these methods.

\begin{table}[ht]
\centering
\renewcommand{\arraystretch}{1.12}
\begin{tabular}{p{3.2cm} p{3.0cm} p{6.8cm}}
\toprule
Method & Error Control & Flagging Behavior \\
\midrule
Tukey's boxplot & None (fixed fence) & Flags a fixed proportion as outliers under normality; the rate increases rapidly with the sample size. \\
Chauvenet-type boxplot & Expected number (PFER) & Flags on average 0.5 outlier per dataset, regardless of the sample size. \\
Holm-type boxplot & Family-wise error rate (FWER) & Controls the probability of at least one false outlier (typically $\leq$5\%). \\
BH-type boxplot & False discovery rate (FDR) & Controls the expected proportion of false outliers among those flagged (typically $\leq$5\%). \\
\bottomrule
\end{tabular}
\caption{
Summary of boxplot outlier detection methods and their error interpretation. PFER: per-family error rate (expected false outlier count); FWER: family-wise error rate (probability of at least one false outlier); FDR: false discovery rate (expected proportion of false outliers among those flagged).
}
\label{tab:boxplot-summary}
\end{table}

Our simulation studies and practical analyses, as illustrated in Section~5 with $n = 31$, reveal that the choice of boxplot method can influence which outliers are detected, even in relatively small samples.
The classical Tukey boxplot and the Chauvenet-type boxplot sometimes yield overlapping but not identical sets of flagged values, especially when the data deviate from normality or exhibit moderate skewness.
For small datasets, both traditional and adaptive boxplot methods are generally reasonable; however, users should consider the distributional characteristics of their data and interpret results with caution.
The Holm-type approach remains particularly appropriate in applications demanding strict control of false positives. For moderate sample sizes, typically 100 to 1000 observations, the Chauvenet-type and BH-type boxplots offer the best balance of sensitivity and specificity.
In large datasets, the fixed-threshold rules should be avoided, as their rate of false positives escalates. The Chauvenet-type boxplot remains robust across all sample sizes, while the BH-type method with a relaxed significance level provides greater sensitivity when multiple outliers are expected.

It is important to recognize, however, that most boxplot-based techniques, including both \texttt{ChauBoxplot} and \texttt{AdaptiveBoxplot}, are developed under the assumption of normality. As illustrated by the real data example in Section~5, real-world datasets such as the provincial Project 985 and Project 211 admission rates can display marked skewness, heavy tails, or other departures from normality due to social, economic, or policy-driven factors. In these cases, even sample-size-adjusted and error-controlled boxplots may produce misleading results. For example, skewed data can result in fences that are either overly strict or overly lenient, which may cause atypical observations to be missed or typical values to be incorrectly labeled as outliers. Practitioners are encouraged to assess distributional characteristics before applying these methods and to consider robust estimation or data transformations as appropriate. Future research directions include developing adaptive boxplot procedures that remain valid under a wider array of data distributions, potentially through transformation-invariant fences or alternative criteria tailored to skewed or heterogeneous datasets.

In summary, our simulation studies and real-world analysis demonstrate that \texttt{ChauBoxplot} and \texttt{AdaptiveBoxplot} provide reproducible, statistically principled tools for modern boxplot-based outlier detection across a wide range of applications. By embedding adaptive error control into familiar graphical displays, these packages empower practitioners to achieve defensible and interpretable results in both exploratory and confirmatory analyses. We recommend these approaches particularly for moderate to large datasets, and emphasize the importance of assessing distributional characteristics to ensure robust and reliable inference.

\section*{Acknowledgement(s)}
The authors sincerely thank the Editor and the two reviewers for their thoughtful and constructive comments, which have greatly improved this article.

\section*{Funding}
Tiejun Tong's research was supported in part by the General Research Fund of Hong Kong (HKBU12300123) and the Initiation Grant for Faculty Niche Research Areas of Hong Kong Baptist University (RC-
FNRA-IG/23-24/SCI/03).
Hongmei Lin's research was supported in part by the National Natural Science Foundation of China (12171310).
Riquan Zhang's research was supported in part by the National Natural Science Foundation of China (12371272, 12531013).

\bibliography{References_boxplots}

% First with apacite package
%\bibliographystyle{apacite}

% Second with natbib package
\bibliographystyle{apalike}

\end{document}